\documentclass[12pt,showpacs,preprintnumbers]{revtex4}

\usepackage{graphicx}
\usepackage{dcolumn}
\usepackage{bm}
\usepackage{amssymb}
\usepackage{epsfig}    
\def\beq{\begin{equation}}
\def\eeq{\end{equation}}
\def\ba{\begin{array}}
\def\ea{\end{array}}
\def\bea{\begin{eqnarray}}
\def\eea{\end{eqnarray}}

\def\End{\end{document}}

\begin{document}                                                              

\title{A modification of the relativistic energy-momentum relation}

\author{Virendra Gupta}
\email{virendra@mda.cinvestav.mx}

\affiliation{
\vspace*{2mm} 
Departamento de F\'{\i}sica Aplicada, 
CINVESTAV-M\'erida, A.P. 73, 97310 M\'erida, Yucat\'an, M\'exico}

\begin{abstract}

A modification of the accepted relativistic energy momentum relation is
suggested. The new relation allows massive particles to have a maximum
velocity $c(m)$ greater than the velocity of light $c$. The effect of the 
 modification suggested here would be most apparent for objects with masses nearly 
equal to the Planck mass.

  
\end{abstract}

\maketitle

    Recently the OPERA collaboration reported the  detection of superluminal neutrinos\cite{opera}.
  Motivated by this we tried to modify the standard energy ,mass and momentum relationship,namely
 
\bea
               E^2  =  m^2 c^4 + p^2 c^2,
\eea
 to
\bea
               E^2  =  m^2 c(m)^4  +  p^2 c(m)^2.
 \eea
        In terms of the velocity $v$,
 
\bea
        E  =   \frac{mc(m)^2}{ \sqrt{1-v^2/c(m)^2}   }  .
\eea

     Here $c(m)$ is a function of a dimensionless variable $\zeta$ which depends on the mass $m$ and is defined below.We
    have simply  replaced $c$ by $c(m)$ in the usual relation.

     Any  modification,in our view, must satisfy the  constraint that for $m=0$, we should have $E=pc$.This means that
     we require $c(0)=c$,so that for $m=0$ we have the usual  energy-momentum relation for electromagnetic radiation.

     For non-relativistic velocities, our modification  gives the Newtonian expression for kinetic energy.
    However,it modifies the rest energy for a massive object. 
  
    The crucial question is what is the variable  $\zeta$   and the function  $c(m)$  .Massive particles have 
     gravitational interaction and it is interesting ,to note that one can define a mass,the Planck mass $ P_M$ in terms of 
    the gravitational constant $G_N $ ,namely
\beq       
       P_M = \sqrt{\hbar c/G_N}\approx 1,22\times 10^{19} GeV/c^2 \approx 2.18\times 10^{-5} gr.
\eeq
       Here $h$=Planck's constant and $c$ is the velocity of light.The dimensionless variable $\zeta$ is 
       defined to be
\beq
           \zeta= m/P_M 
\eeq
       We define $c(m)$  to be of the form
\beq
             c(m)=c[ 1+F(\zeta)]
\eeq
          where $F$ is a positive analytic function of  $\zeta$ with $F(0)=0$ .Positivity of $F$ is necessary for
          matter to have superluminal velocities.From the definition it is clear that $\zeta$ is extremely small
         for sub-atomic particles (electron ,proton etc.) and is very large for ordinary masses.For example,
           for $m=20 gr$,  $\zeta\sim 10^6$ and enormous for ordinary and stellar  objects.Since,the value of
           $ F(\zeta)$ will affect the rest mass of the object,one has to choose it carefully.From the
           practical point of view,we consider

   \beq     
                   F(\zeta)= \zeta^n exp(-\zeta^n),
 \eeq
           where  $n$ is a positive real number. This function has a maximum value $e^{-1} =0.367879..$ for $\zeta=1$ 
           independent of the   value of $n$. It is centered around $\zeta=1$. As $n$ increases the height remains 
           the same, but it becomes narrower and narrower. For extremely large $n$ (tending to infinity), this sequence of
          functions tends to a vertical line of height $e^{-1} $ located at $\zeta=1$.In this limit, it is like a 
          "finite Dirac $\delta$ function" only non-zero for $\zeta=1$. The function is plotted in Fig.1 for
           $n=50,100$~and~$1000$. One can see clearly how the function shrinks rapidly as $n$ increases and will
           be practically narrow as a  straight line located at $\zeta=1$ for much  larger $n$.

          Thus, for extremely large $n$, the value of the  usual rest  mass $mc^2$ will not be affected for bodies 
           which are very light(subatomic particles) or very  heavy (eg.planets)  compared to the Planck mass, but will 
          affect those  with masses of the order of the Planck mass. Such a choice for $n$ is necessary, since our particle 
          accelerators work and we understand planetary motion. These provide extremely stringent tests for the usual 
          energy-momentum relation. However, we do not have  stringent tests for masses of the order
          of the  Planck mass. It would be interesting to have experimental tests of the energy-momentum relation
          for this mass range.

\begin{figure}
\includegraphics[width=9.0cm,height=7.5cm]{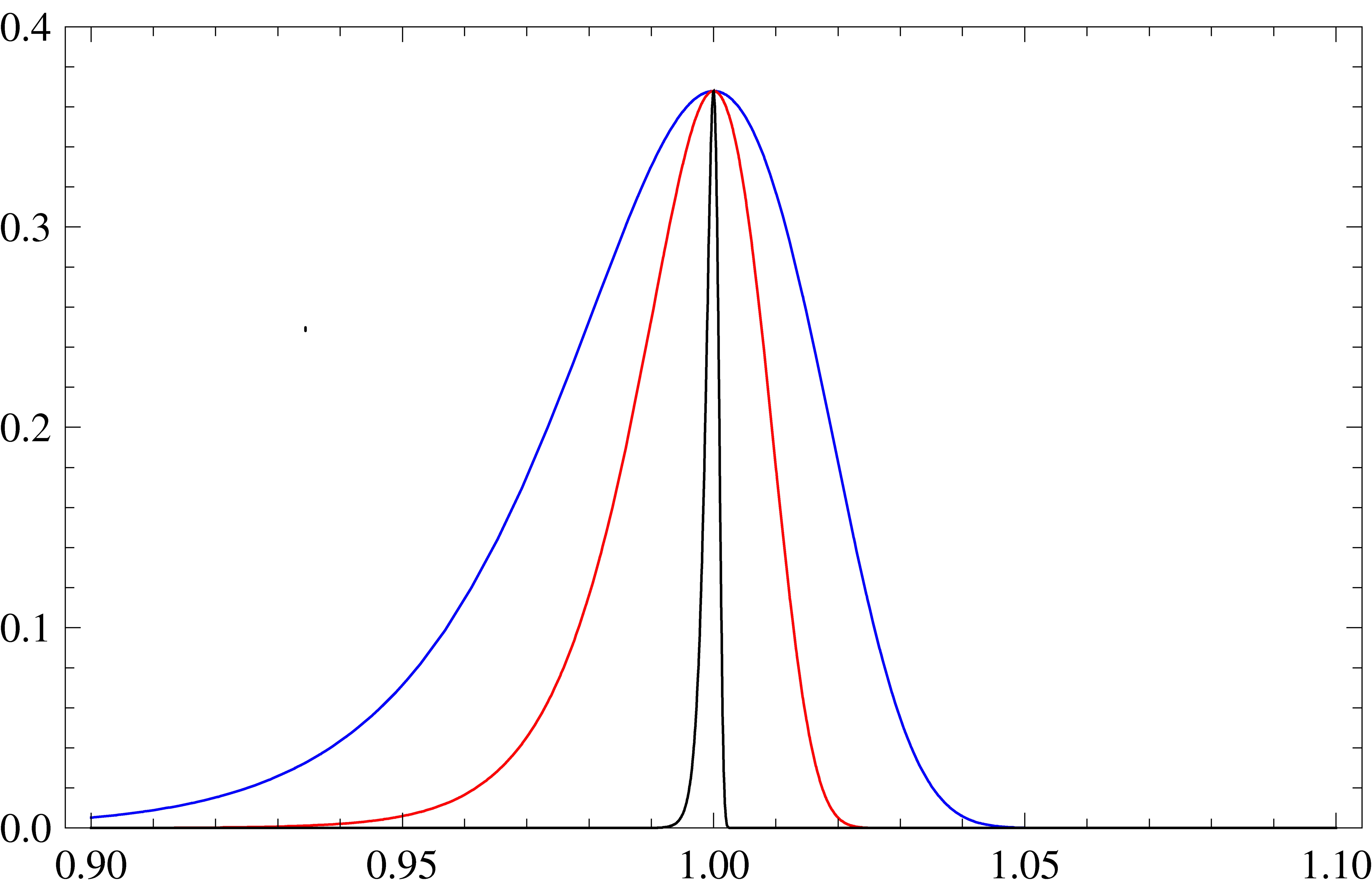}
\caption{The function $F(x)=x^n e^{-x^n}$ ($x\equiv \zeta$)
for $n=50,100,1000$ in
color blue, red and black respectively. The function narrows rapidly with increasing $n$ .
The maximum height at $x=1$ is independent of $n$. }
\label{figure1}

\end{figure}

\noindent
{\bf Acknowledgments}~~~It is a pleasure to thank Francisco Larios for his unstinting kind help in the preparation 
of this manuscript.I would like to thank him and Antonio Bouzas for discussions. 
  I would also 
like to  thank Conacyt and SNI for support.

\end{document}